\documentclass{article}

\usepackage[english]{babel}
\usepackage[official]{eurosym}
\usepackage{multirow}
\usepackage[letterpaper,top=2cm,bottom=2cm,left=3cm,right=3cm,marginparwidth=1.75cm]{geometry}

\usepackage{amsmath}
\usepackage{graphicx}
\usepackage[colorlinks=true, allcolors=blue]{hyperref}
\usepackage{natbib}
\usepackage{lscape}
\usepackage{threeparttable}

\title{Do You Trust ChatGPT? -- Perceived Credibility of Human and AI-Generated Content}
\author{Martin Huschens$^1$, Martin Briesch$^2$, Dominik Sobania$^2$, Franz Rothlauf$^2$}

\date{$^1$University of Applied Sciences Mainz, Germany \\
{\small \texttt{martin.huschens@hs-mainz.de}}\\
$^2$Johannes Gutenberg University Mainz, Germany\\
{\small \texttt{\{briesch,dsobania,rothlauf\}@uni-mainz.de}}}

\begin{document}
\maketitle

\begin{abstract}
This paper examines how individuals perceive the credibility of content originating from human authors versus content generated by large language models, like the GPT language model family that powers ChatGPT, in different user interface versions. Surprisingly, our results demonstrate that regardless of the user interface presentation, participants tend to attribute similar levels of credibility. While participants also do not report any different perceptions of competence and trustworthiness between human and AI-generated content, they rate AI-generated content as being clearer and more engaging. The findings from this study serve as a call for a more discerning approach to evaluating information sources, encouraging users to exercise caution and critical thinking when engaging with content generated by AI systems. 
\end{abstract}

\section{Introduction}

In an era where internet-based digital applications play an indispensable role in finding and confirming information, the credibility of content as well as sources has become a critical concern \citep{Luo2022, Winter2014, Kiousis2001, Flanagin2000}. With the advent of online encyclopedias in the early 2000 it was heavily discussed whether Wikipedia can be acknowledged as a trusted and credible source. Only years later in 2005 a Nature article showed that Wikipedia was not significantly less accurate than Encyclopedia Britannica \citep{Giles2005}. Nowadays generative artificial intelligence (AI) applications based on large language models (LLMs) such as ChatGPT, Google Bard, Claude 2 or Bing Chat and others have revolutionized the landscape of the internet. Internet users turn to this new generation of applications to retrieve information due to their incomparable ease of use \citep{Xu2023}. However, the main difference between Wikipedia and generative AI is that Wikipedia presents human-generated and curated content while generative AI creates content without human intervention and is known to generate erroneous output \citep{borji2023categorical}. The origin of the content, however, may be difficult for humans to distinguish, which poses several potential risks.

Therefore, this paper studies credibility perceptions of human-generated and computer-generated content in different user interface (UI) settings. We conduct an extensive online survey with overall 606 English speaking participants and ask for their perceived credibility of text excerpts in different UI settings (ChatGPT UI, Raw Text UI, Wikipedia UI) while also manipulating the origin of the text: either human-generated or generated by an LLM (``LLM-generated'').

Surprisingly, our results demonstrate that regardless of the UI presentation, participants tend to attribute similar levels of credibility to the content. Furthermore, our study reveals an unsettling finding: participants perceive LLM-generated content as clearer and more engaging while on the other hand they are not identifying any differences with regards to message's competence and trustworthiness.

This finding highlights a pressing concern. Despite the extensive training and continuous advancement in AI-driven content generation, the potential for errors, misperceptions, and even hallucinatory behavior persists within these applications \citep{jang2023consistency, shen2023chatgpt}. Hence, the public's increasing reliance on these platforms for information retrieval poses significant risks. As our study shows, users seem to consume content without distinguishing its origin. Thus, the potential for the spread of misinformation and the erosion of critical thinking skills cannot be overlooked. This is especially disturbing when thinking about long-term effects. As the accessibility and adoption of AI-generated content continue to expand, society faces a crucial juncture. Users must approach information consumption with a higher awareness of the limitations and inherent biases embedded within AI-generated content. 

Following this introduction, Section \ref{sec:background} provides the theoretical background to our study as well as literature reviews on message credibility and LLMs. In Section \ref{sec:method}, we report method and study design of the questionnaire. Section~\ref{sec:results} presents our results on participant's credibility perceptions followed by a discussion of the implications of our study (Section~\ref{sec:discussion}). In Section~\ref{sec:limitations}, we address potential limitations before concluding the paper in Section~\ref{sec:conclusion}.

\section{Theoretical Background}
\label{sec:background}
\subsection{Credibility Perceptions}

Assessing  whether an information source or message is trustful or credible plays an important role in the interpretation and assimilation of information. Research on credibility perceptions has a long-standing tradition and has its roots already in the 1950s. Credibility in general is defined as ``[\dots] a perceptual state, i.e., the outcome of an attribution process in which recipients of messages form judgments about their sources and therefore assess them as credible or not.'' \citep{Jackob2008}. Credibility research in general considers three distinct levels: (1) source credibility, (2) media credibility and (3) message credibility \citep{Kiousis2001, Appelman2016, Winter2014}. Put differently: When making personal judgements whether a piece of information is trustful or credible, we take three aspects into account: (1) who is the sender of the information (\textit{source credibility}); (2) through which channel is the information presented (\textit{media credibility}) and (3) how is the message formulated (\textit{message credibility}) \citep{Hellmueller2012}.

Early contributions in the research area of credibility perceptions focus on the assessment of information sources. This research stream refers to certain communicator characteristics and studies how these influence the perception of messages \citep{Kiousis2001}. In most studies, the communicator is defined as either an individual, a group, or an organization. Notably, \cite{Hovland1951} laid the groundwork for understanding source credibility. They demonstrated that the same content presented by two different sources (a well-known and credible expert vs.~an untrustworthy source) was perceived on different ends of a credibility continuum. Furthermore, they also studied the development of credibility perceptions over time and coined the term of the ``sleeper effect'': since persons forget about the information source, messages from untrustworthy sources become more credible over time \citep{Hovland1951}. Already back then credibility was found to be a multi-dimensional construct with (at least) two dimensions: perceived \textit{expertness (competence)} and perceived \textit{trustworthiness} \citep{Hovland1953, Winter2014, Berlo1969}. Perceived \textit{competence} relates to the perception whether a source is able to provide valid statements on a topic, while perceived \textit{trustworthiness} refers to the perception whether a source is willing to communicate the correct information \citep{Winter2014}. Later conceptualizations of source credibility extended these communicator-related dimensions by, e.g., safety, qualification, expertise or dynamism \citep{Berlo1969, Whitehead1968}.

In addition to that, credibility perceptions may also be influenced by non-source factors, such as the medium or channel of delivery. This is why media credibility research has focused more on the channel through which content is delivered rather than the sender (or senders) of that content \citep{Kiousis2001}. In times of diverse media channels this research area is of special interest: credibility extends beyond individual sources to encompass the credibility of the medium itself.  Research in this area explores differential perceptions of message's credibility, e.g., television, radio, newspapers, or online news (so called \textit{channel credibility competition}). Again,  credibility is  conceptualized as a multi-dimensional construct formed by several sub-constructs such as believability, accuracy, trustworthiness, bias, and completeness of information \citep{Flanagin2000}. More recently, media credibility is increasingly studied in online contexts, e.g., identifying relevant site design features to build credibility \citep{Lowry2014, Fogg2001, Flanagin2007}, student's ability to evaluate information on the internet \citep{McGrew2018}, a person's detection accuracy of fake and real news on social media \citep{Luo2022}, or also credibility perceptions of GPT-2 generated texts compared to New York Times, Huffington Post or Fox News articles \citep{kreps2022all}.

Besides the source, the credibility of a message also depends on the message characteristics itself. Thus, another line of research, namely message credibility, focuses on characteristics of messages that
could make a message more credible. \cite{Appelman2016} defined message credibility as ``[\dots] an individual’s judgment of the veracity of the content of communication.'' While source and media credibility research often refer to trust in the communicator or the media as a whole, message credibility refers more narrowly to a message, event or piece of information \citep{Dalen2020}. Following this description, the object of message credibility is more narrow than the object of trust. Again, message credibility is perceived as a multi-dimensional construct and  frequently operationalized using sub-constructs like trustworthiness, accuracy, fairness, balance, absence of opinion, completeness, authenticity and believability \citep{Sundar1999}. \cite{Appelman2016} argue to measure message credibility by asking participants to rate how well certain adjectives describe content, namely \textit{accurate}, \textit{authentic}, and \textit{believable}. In sum, they suggest ten indicators to capture important aspects of message credibility (\textit{complete, concise, consistent, well-presented, objective, representative, no spin, expert, will have impact, professional}, \cite{Appelman2016}).

\subsection{Generative AI and LLMs}

LLMs are an emerging technology based on deep learning approaches, specifically transformer models \citep{vaswani2017attention} like GPT \citep{brown2020language}. These LLMs achieve highly competitive results in many natural language processing tasks, like machine translation \citep{brown2020language}, source code generation \citep{chen2021evaluating,sobania2022choose}, or automated program repair \citep{sobania2023ananalysis}, and (will) influence the life of many people. Applications like ChatGPT even became the software with the fastest growing user base in human history so far.\footnote{\url{https://www.zdnet.com/article/chatgpt-just-became-the-fastest-growing-app-of-all-time/}} These well-known software products are based on generative auto-regressive models that are trained in a multi-stage process \citep{ouyang2022training}. First, the model is pre-trained in an self-supervised manner on a large and unstructured dataset by predicting the next token in a sequence. Put simply, in this first step, statistical relationships between text components (e.g., characters, words, sentences, etc.) are learned. This way, a high-quality sentence completion model can be build. In the second step, such models are fine-tuned in a supervised manner in order to achieve even better results in special application domains, such as summarization or question answering. Specifically, a user input (also known as \textit{prompt}) is given and the model has to complete this input with a desired output specified by a human. However, to enable applications such as ChatGPT or Bard, one more step is necessary. For further improvement, reinforcement learning by human feedback is incorporated. Specifically, this means that a reward model is first trained using humans labeling LLM-generated answers. These human ``labeler'' rank multiple answers generated by the LLM in descending order regarding quality. These rankings are then used to train the reward model which then can be used as a reward function for a reinforcement learning approach (e.g., proximal policy optimization). This reinforcement learning approach further improves the model's quality and helps to align the model with the user's intent \citep{ouyang2022training}.

However, as LLMs are mainly pre-trained on uncurated data available on the internet, including error-prone, fictitious and biased data, their output should be treated with caution. Even if attempts are made to align the underlying language model through fine-tuning and reinforcement learning by human feedback, the problematic data from the underlying pre-trained model may still emerge. Even some OpenAI researchers stated that their ``[\dots] models are neither fully aligned nor fully safe; they still generate toxic or biased outputs, make up facts, and generate sexual and violent content [\dots]'' \citep{ouyang2022training}. It is also well documented, that LLM-based applications like ChatGPT are prone to errors when used in practice \citep{borji2023categorical}. 
Overall, scientists and politicians are concerned about generative AI and many relevant researchers have signed a statement that generative AI should be treated as a societal risk on the level of nuclear war and pandemics.\footnote{\url{https://www.safe.ai/statement-on-ai-risk}}

Therefore, the question arises whether users perceive LLM-generated texts differently from texts from trustworthy sources or whether they cannot distinguish between them. The later could amplify the dangers posed by LLMs to society (such as the automatic generation of fake news \citep{kreps2022all} etc.). Consequently, in this work we investigate the perceived credibility of LLM-generated textual content compared to content generated and curated by humans.

\section{Method}
\label{sec:method}
\subsection{Survey Design}
\begin{figure}[ht!]
	\centering
	\begin{tabular}{@{}c@{}}
	    \small (a) ChatGPT UI {\smallskip}\\\textbf{}
		\frame{\includegraphics[width=0.54\textwidth]{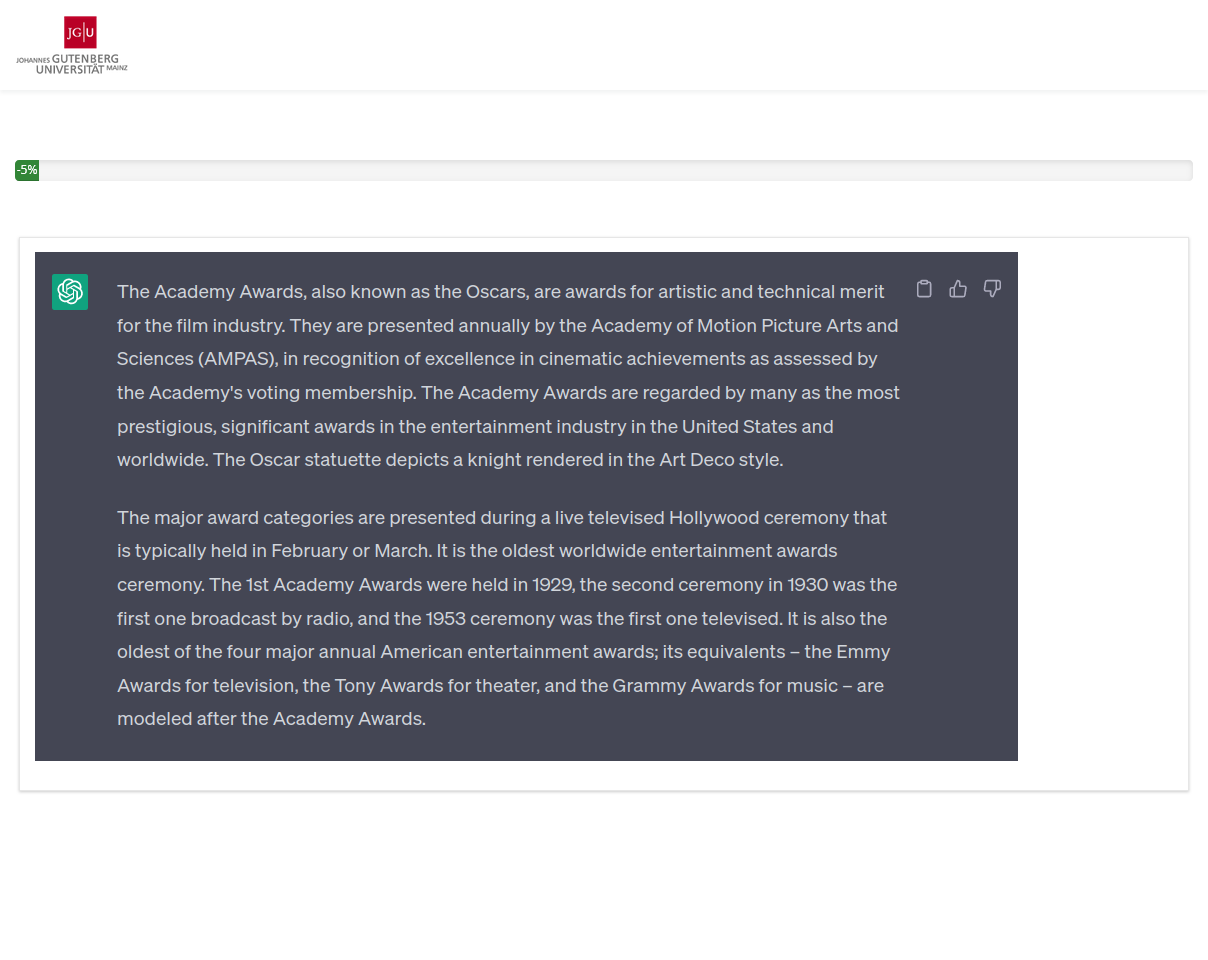}}	\\	
		\small(b) Raw UI {\smallskip}\\
		\frame{\includegraphics[width=0.54\textwidth]{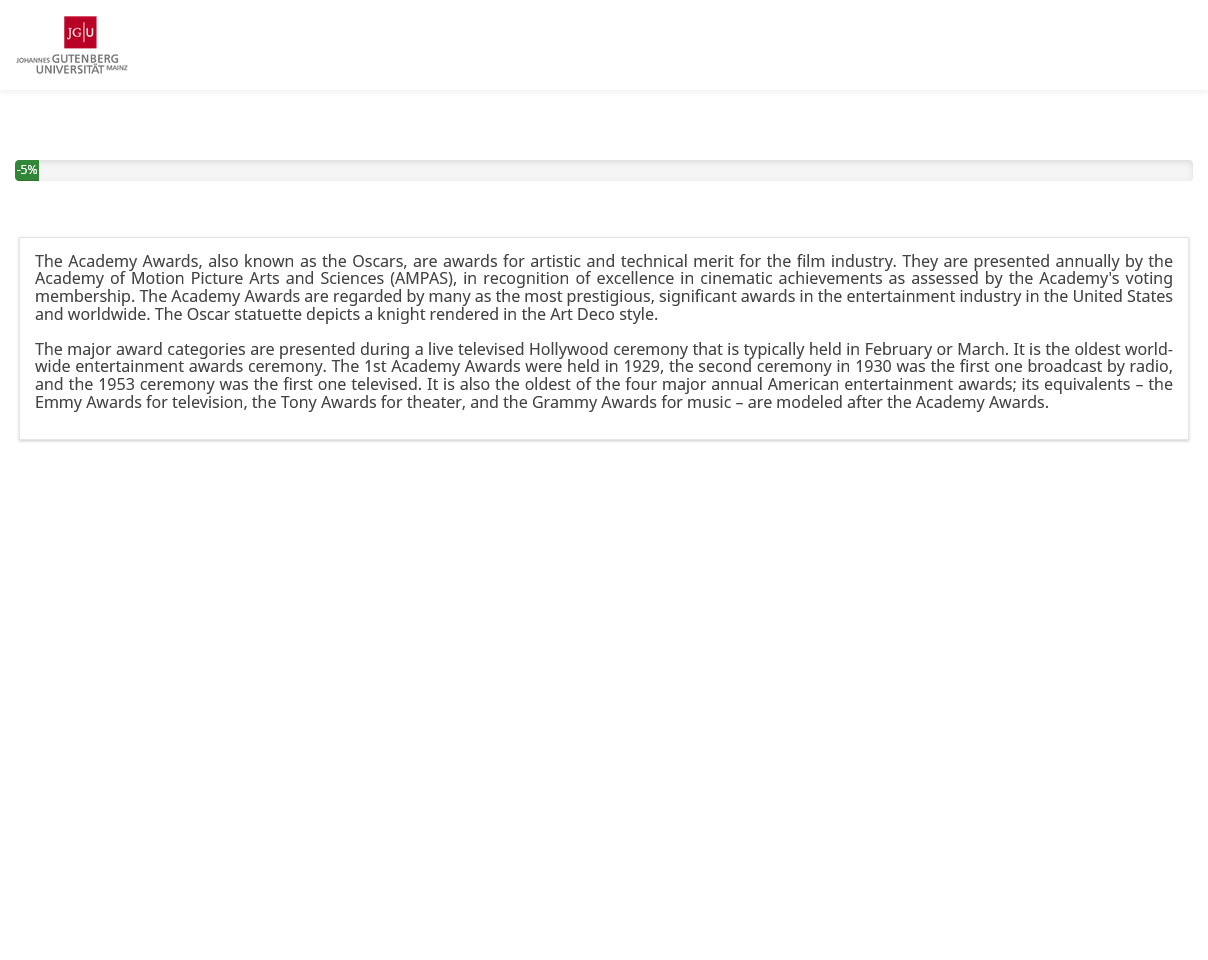}}\\
		\small (c) Wikipedia UI {\smallskip}\\
		\frame{\includegraphics[width=0.54\textwidth]{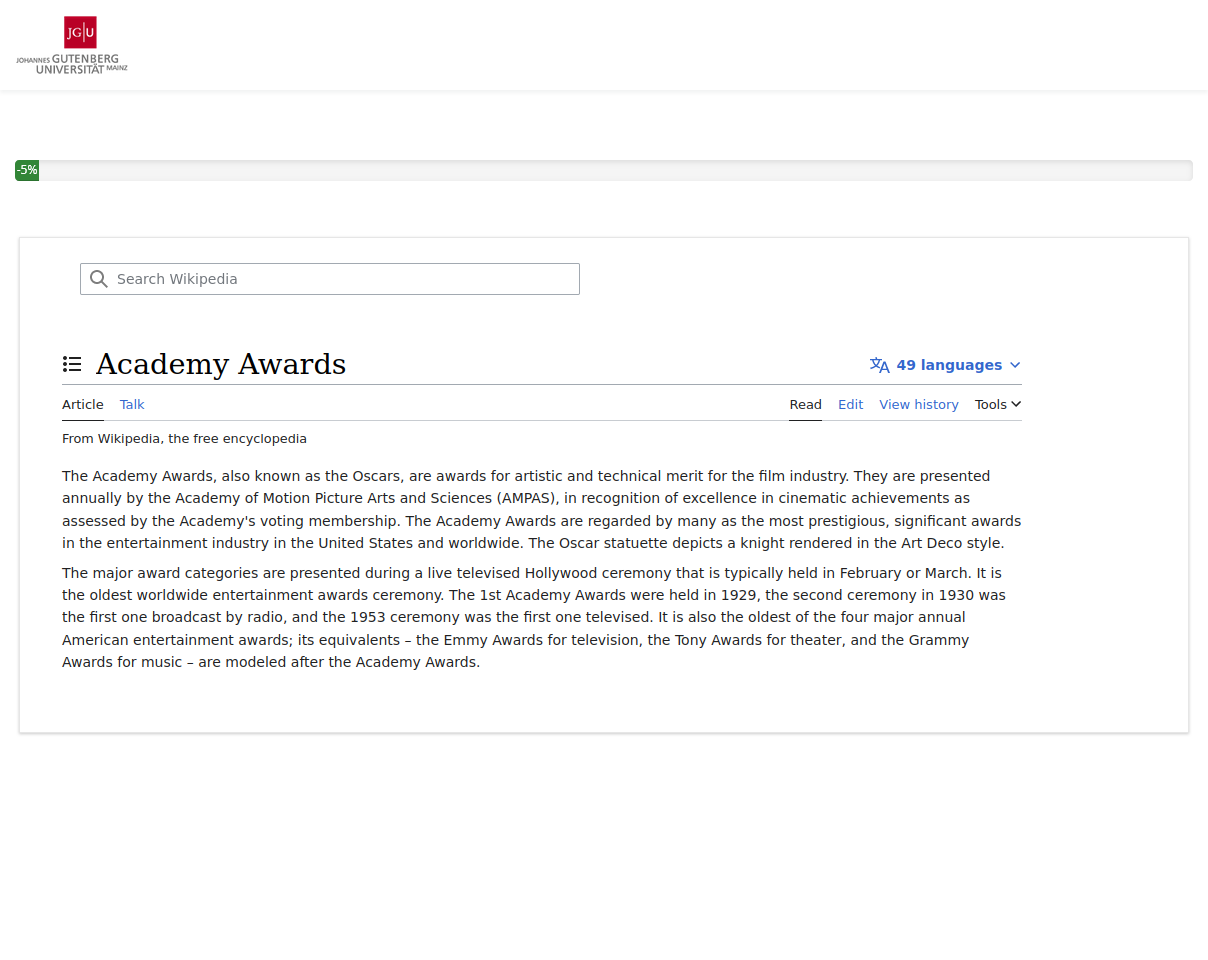}}\\
	\end{tabular}
	\caption{Treatment Conditions - UI Versions, Topic of Example Text: Academy Awards, human-generated text version}
	\label{fig1:treatments}
\end{figure}

To investigate the influence of different UI conditions on users' perceived credibility of information, our research employs a survey approach. The study's questionnaire-based format presents each participant with a set of four images, each containing a concise text excerpt displayed under one of three distinct UI conditions: ChatGPT UI, Raw Text UI, or Wikipedia UI (depicted in Figure \ref{fig1:treatments}, Treatment Conditions). Random allocation places each participant into one of these UI treatment groups. After engaging with each text excerpt, participants had to answer two comprehension questions that served as attention checks. The exact comprehension questions can be found in Appendix \ref{Appendix:attention_check}. After that they participants are asked to respond to a series of 11 items crafted to assess message credibility (detailed in Section \ref{sec:measure}).

The text excerpts cover the following four topics: Academy Awards, Canada, Malware and US Senate (see Figure \ref{fig1:treatments} for the text example on the Academy Awards). These topics are drawn from the top-100 Wikipedia articles between December 1, 2007, and January 1, 2023, ensuring their relevance and prominence. Participants see these texts in a deterministic order. 

For each topic we created two text versions: a human-generated version and an LLM-generated version. 
The human-generated version is drawn from Wikipedia, specifically, we took the introductory part of the article and removed all hyperlinks and references from the text.
The LLM-generated version has been generated using ChatGPT.\footnote{We used the ChatGPT (GPT-3.5) version from 23 March 2023.} We prompted ChatGPT using the query: ``\texttt{Write a dictionary article on the topic "[TITLE]". The article should have about [WORDS] words}''. \texttt{[TITLE]} is the respective topic title and \texttt{[WORDS]} is the number of words in the corresponding human-generated text. This resulted in LLM-generated content with roughly the same number of words as its human-generated Wikipedia counterpart.

Furthermore, we manually checked the LLM-generated texts for factual errors and did not find any major mistakes.
The exact text excerpts can be found in Appendix~\ref{Appendix:Text}.

For each of the four topics, participants see either a human-generated or a LLM-generated text in random order, and each participant is presented with exactly two human-generated and two LLM-generated texts, one for each topic. This arrangement enables the investigation of the combined effects of UI conditions and text sources on participants' perceived perceptions of message credibility.

In order to ensure the validity and reliability of our study's findings, we also gather information on relevant covariates that could potentially influence the observed outcomes (see Table \ref{tab:controls} for the descriptives). As shown in Table \ref{tab:controls}, according to the results of the Kruskal-Wallis tests, one-way Analysis of Variance (ANOVA) and Pearson’s Chi-Square tests, no significant difference can be detected across treatments in gender, age, education, income, English language proficiency, spoken language, knowledge of LLMs, usage of LLMs or usage of online encyclopedias. This supports that we were successful in assigning the subjects to the treatment groups and avoids any confoundings based on these important covariates. This means we can rule out any structural group differences as being the cause of differences found in credibility perceptions between groups. 
\newpage

	\begin{table}[ht!]
		\centering
		\caption{Controls - Descriptives and Randomization Check}
		\footnotesize
		\label{tab:controls}
		\begin{threeparttable}
			\begin{tabular}[]{@{}lccccccc@{}}
				\hline\noalign{\smallskip}
				& ChatGPT & Raw  & Wikipedia & All & ANOVA & KW & \\
				\hline\noalign{\smallskip}
				\textbf{N} & 201 & 210 & 195 &  606 &-&-&\\
				\textbf{Gender} &  &  &   &  & 0.965\textsuperscript{n.s., a}&\\
				\> Male & 94 & 104 &   88 & 286 &&&\\
				\> Female & 102 & 103 &  103 & 308 &&&\\
				\> Other & 5 & 5 &  4 & 14 &&&\\
				\textbf{Age}&   &  &  &   & 0.474\textsuperscript{n.s.}& 1.527\textsuperscript{n.s.}&\\
				\> Mean & 37 & 37.81 &  38.35 & 37.71 &&&\\
				\> Median & 34 & 35 &  38 & 35.50 &&&\\
				\> SD & 14.08 &  14.57 &  13.22 & 13.96 &&&\\
				\> Min & 14 & 15 & 14 &  14 &&&\\
				\> Max & 95 & 95 & 70 &  95 &&&\\
				\textbf{Education}& &  &  & &   5.022\textsuperscript{n.s., a}&-&\\
				\> Less than high school degree & 1 & 2 &  3 & 6 &&&\\
				\> High school degree or equivalent (e.g., GED) & 30 & 30 &  30 & 90 &&&\\
				\> Some college but no degree & 49 & 48 &  44 & 141 &&&\\
				\> Associate degree & 20 & 18 &  10 & 48 &&&\\
				\> Bachelor degree & 70 & 79 &  74 & 223 &&&\\
				\> Graduate degree & 31 & 33 &  34 & 98 &&&\\
				\textbf{Income} & &  &  &   & 17.015\textsuperscript{n.s., a}&-&\\
				\> \$250,00   or less & 36 & 36 &  27 & 99 &&&\\
                \> \$250,01 to \$500,00   & 11 & 11 &  15 & 37 &&&\\
                \> \$500,01 to \$750,00   & 10 & 15 &  10 & 35 &&&\\
                \> \$750,01 to \$1.000,00   & 11 & 11 &  7 & 29&&& \\
                \> \$1.000,01 to \$1.500,00  & 17 & 15 &  14 & 46&&& \\
                \> \$1.500,01 to \$2.000,00   & 16 & 13 &  18 & 47 &&&\\
                \> \$2.000,01 to \$2.500,00   & 22 & 16 &  12 & 50 &&&\\
                \> \$2.500,01 to \$3.000,00   & 20 & 23 &  16 & 59 &&&\\
                \> \$3.000,01 or more  & 55 & 69 &  76 & 200 &&&\\
				\textbf{English Proficiency}  & &  &  & & 1.565\textsuperscript{n.s.}& 1.870\textsuperscript{n.s.}&\\
				\> Mean & 4.97 & 4.97 &  4.93 & 4.96 &&&\\
				\> Median & 5.00  & 5.00 &  5.00 & 5.00 &&&\\
				\> SD & 0.17 & 0.16 &  0.35 & 0.23 &&&\\
				\> Min & 1.00 & 1.00 & 1.00  & 1.00 &&&\\
				\> Max & 5.00 & 5.00 & 5.00  & 5.00 &&&\\
				\textbf{Spoken Language}  & &  &  & & 0.002\textsuperscript{n.s., a}&-&\\
			    \> English & 199 & 208 &  192 & 599 &&&\\
			    \> Spanish & 1 & 2 & 2  & 5 &&&\\
			    \> Other & 1 & 0 & 1  & 2 &&&\\
				\textbf{Knowledge LLM\textsuperscript{*}} &&  &  &  &  0.673\textsuperscript{n.s.} & 1.311\textsuperscript{n.s.}& \\
				\> Mean &  2.93 & 2.87 &  2.80 & 2.87&&& \\
				\> Median & 3.00 & 3.00 &  3.00 & 3.00&&& \\
				\> SD & 1.06 & 1.11 &  1.05 & 1.07&&& \\
				\> Min & 1.00 & 1.00 &  1.00 & 1.00&&& \\
				\> Max & 5.00 & 5.00 &  5.00 & 5.00&&& \\
				\textbf{Usage LLM\textsuperscript{**}} &&  &  &   & 12.996\textsuperscript{n.s., a}&-& \\
			    \> Every Day & 10 & 9 &  14 & 33&& \\
				\> Several times a week & 45 & 53 &  34 & 132&&& \\
				\> Once a week & 23 & 28 &  29 & 80&&& \\
				\> Once a month & 32 & 21 &  21 & 74&&& \\
				\> Less often & 43 & 34 &  36 & 113&&& \\
				\> Never & 44 & 58 &  55 & 157&&& \\
				\> I don't know any of these applications &4& 7 & 6 &  17 & && \\
				\textbf{Usage Online Encyclopedias\textsuperscript{***}} &&  &  &   & 11.174\textsuperscript{n.s., a}&-& \\
				\> Every Day & 25 & 18 &  14 & 57&&& \\
				\> Several times a week & 81 & 89 &  77 & 247&&& \\
				\> Once a week & 39 & 44 &  44 & 127&&& \\
				\> Once a month & 31 & 21 &  27 & 79&&& \\
				\> Less often & 21 & 34 &  27 & 82&&& \\
				\> Never & 4 & 4 &  5 & 13&&& \\
				\> I don't know any of online encyclopedia & 0 & 0 &  1 & 1&&& \\

				\hline\noalign{\smallskip}
			\end{tabular}
			\begin{tablenotes}
				\item \scriptsize \textit{Note.} ANOVA = F-Statistic of Analysis of Variance for group difference between ChatGPT UI, Raw Text UI and Wikipedia UI. KW = H-Statistic of Kruskal-Wallis Test for group difference between ChatGPT UI, Raw Text UI and Wikipedia UI. \textsuperscript{a} = Results of a Pearson's Chi-Square test. \textsuperscript{n.s.} = not significant. * Knowledge LLM = ``How would you rate your knowledge regarding large language models such as ChatGPT, Bard, LlaMa or similar on a scale from 1-5, where 1 is non-existent and 5 is expert-level.'', ** Usage LLM =  ``How often do you use applications of large language models such as ChatGPT, Bing Chat, Bard, LlaMa or similar?'', *** Usage Online Encyclopedias = ``How often do you use online encyclopedias such as Wikipedia or similar?''
			\end{tablenotes}
		\end{threeparttable}
	\end{table}

\subsection{Participants}
To ensure a methodical and well-regulated recruitment of participants for our study, we used the online platform Prolific.\footnote{\url{https://www.prolific.co}} This platform offers a diverse and extensive pool of potential participants, facilitating efficient data collection while giving us some control over participant demographics. In alignment with the findings of \cite{Peer2017}, we favored Prolific over Amazon mTurk due to data quality advantages. We recruited our sample over the period of two weeks in July 2023. Sample descriptives can be found in Table \ref{tab:controls}.

Our study's recruitment efforts were exclusively centered around English-speaking participants who accessed the study via desktop devices. This approach aimed to eliminate potential confounding factors tied to language proficiency or display dimensions, thereby upholding the uniform readability of the presented text excerpts over all UI conditions. In preparation for our study, we conducted an a priori power analysis\footnote{For conducting the power analysis we used G*Power 3.1.9.7 for Windows, 20 March 2020, \citep{Faul2009}.} to ascertain the minimal sample size necessary for detecting a small population effect size \citep{Faul2009, Raudenbush2000}. This analysis indicated a requirement ranging from 403 to 480 participants \citep{Cohen1988}. Ultimately, our sample yielded complete answers from a total of 606 participants, surpassing the established sample size threshold. Participants who successfully completed the study were compensated with a payment of \pounds 7.50. The compensation not only serves as an incentive for participation but also helps maintain participants' engagement and motivation throughout the study. Our attention checks proved that participants read the texts carefully (see the very low error rates in Table \ref{tab:results_kw}). The median processing time required for participants to complete the entire questionnaire was found to be 12:48 minutes.

\subsection{Measure Development and Validation}
\label{sec:measure}

 Our reflective measurement model of message credibility was inspired by well-established questionnaires and measurement approaches in previous studies \citep{Flanagin2007, Winter2014, Berlo1969, Appelman2016}. Following these approaches, we operationalized credibility perceptions with four reflective factors: \textit{competence, trustworthiness, clarity and engagement}. While the first two sub-constructs (\textit{competence} and \textit{trustworthiness}) measure source credibility following early contributions by \cite{Hovland1951}\footnote{Another often employed sub-construct is called ``Dynamism'' which has been introduced by \cite{Berlo1969}. They define it as the extent to which a source is ``“fast, energetic [\dots] bold [\dots] colorful, and confident'' \citep[p.~567]{Berlo1969}. In online contexts this sub dimension of credibility is often framed as the way in which information is presented (i.e., graphic designs and aesthetics; \cite{Robins2008}. We deliberately excluded this sub-construct because we were not interested in graphical designs but rather in the message and source perceptions itself}, the latter two (\textit{clarity} and \textit{engagement}) focus on the quality of the message itself (message credibility). Based on these factors, participants had to express their perceptions of four texts by rating 11 items on 5-point Likert scale. They were given the following task: ``Rate the text according to the following adjectives from 1 = describes very poorly to 5 = describes very well.'' Since we used only selected sub-scales, developed some new items, and combined them into a new measurement instrument, we conducted a pre-study with 60 participants and did a confirmatory factor analysis (CFA) to confirm the model fit of our four-factor measurement model as well as its reliability and validity. Since the results of the pre-study showed several cross-loadings of items as well as a non-satisfactory model fit we deleted some items and re-phrased others. The final measurement model can be assessed in Table \ref{tab:cfa}.
 
 \begin{table}[ht!]
	\caption{Confirmatory Factor Analysis - Message Credibility}
	\centering
	\label{tab:cfa}
	\footnotesize 
	\begin{threeparttable}
    	\begin{tabular}{lccc}
    		\hline\noalign{\smallskip}
    		Factor and Item & Std. Loading & CR & Cronbach's $\alpha$ \\
    		\hline\noalign{\smallskip}
    		 \textbf{Competence} &&0.81           & 0.91\\
    		 \> accurate & 0.818 && \\[0,5ex]
    		 \> complete & 0.725 && \\[0,5ex]
    		 \> knowledgeable  & 0.805&& \\[0,5ex]
    		 \textbf{Trustworthiness} && 0.90& 0.84\\
    		 \> honest & 0.848&& \\[0,5ex]
    		 \> trustworthy & 0.892&& \\[0,5ex]
    		 \> reliable & 0.866&& \\[0,5ex]
    		 \textbf{Clarity} && 0.74& 0.93\\
    		 \> clear & 0.862&& \\[0,5ex]
    		 \> confusing\textsuperscript{*} & 0.536&& \\[0,5ex]
    		 \> understandable &0.826&& \\[0,5ex]
    		 \textbf{Engagement} && 0.89& 0.79\\
    		 \> interesting &0.875&& \\[0,5ex]
    		 \> maintaining attention &0.916&& \\[0,5ex]
    		\hline 
    		\end{tabular}
    	\begin{tablenotes}
				\item \scriptsize \textit{Note.} N = 606. Model Fit Indices: $\chi\textsuperscript{2}$ = 16,300.471; d.f. = 55 ; CFI = 0.976; TLI = 0.966; NNFI =  0.966; RMSEA =  0.065; SRMR = 0.035. CR = composite reliability. \textsuperscript{*} = Reverse coded item. Items measured on a 5-point Likert-Scale: ``Rate the text according to the following adjectives from 1 = describes very poorly to 5 = describes very well.''
			\end{tablenotes}
	\end{threeparttable}
\end{table}

Table \ref{tab:cfa} shows the items used and the respective standardized factor loadings. All items but one (Clarity - ``confusing'': $\lambda = .536$) loaded high with standardized coefficients ranging from 0.536 to 0.916 and significant ($p<.001$) on their parent factor and, hence, met the suggested threshold of indicator reliability of .700 \citep{Hair2011}. Nevertheless, we kept the item ``confusing'' in our measurement model: ``[I]ndicators with loadings between 0.40 and 0.70 should only be considered for removal from the scale if deleting this indicator leads to an increase in composite reliability above the suggested threshold value'' \citep[][p.145]{Hair2011} which was not the case in our analysis since all composite reliability scores already met the suggested threshold of .700 (\cite{Nunnally1994}, see Table \ref{tab:cfa}, column CR).

\begin{table}[ht!]
	\caption{Correlation Between Factors -- Discriminant Validity}
	\centering
	\label{tab:discriminant_validity}
	\footnotesize
	\begin{threeparttable}
		\begin{tabular}{lcccc}
			\hline\noalign{\smallskip}
			&Competence&Trustworthiness&Clarity&Engagement\\
			\hline\noalign{\smallskip}
			Competence&\textbf{.775}&&&\\[0,5ex]
			Trustworthiness&.681  &\textbf{.869}&&\\[0,5ex]
			Clarity&.643  &.596&\textbf{.704}&\\[0,5ex]
			Engagement&.539  &.520&.486&\textbf{.894}\\[0,5ex]
			\hline\noalign{\smallskip}
		\end{tabular}
		\begin{tablenotes}
			\item \scriptsize \textit{Note.} Diagonal elements are the square root of Average Variance Extracted (AVE) per factor. Off-diagonal elements are the inter-factor correlations. 
		\end{tablenotes}
	\end{threeparttable}
\end{table}

The CFA revealed that the model fit was satisfactory\footnote{The model fit was assessed using the R package ``lavaan'' version 0.5-16 in R version 4.3.1\citep{Rosseel2012}.} (Robust CFI = 0.967, Robust TLI = 0.966, NNFI = 0.966, RMSEA =  0.065, SRMR = 0.035). For construct validity, we examined both convergent and discriminant validity \citep{Gefen2005, Cronbach1955}. First, the square root average variance extracted (AVE) for all factors was above 0.500 which confirmed convergent validity of our measurement instrument (see diagonal elements in Table \ref{tab:discriminant_validity}). Second, discriminant validity was tested by comparing the square root of the AVE and the inter-construct correlations \citep{Fornell1981}. Table \ref{tab:discriminant_validity} shows that, for all cases, the correlation between two constructs (off-diagonal values) was lower than the square root of the AVE (diagonal elements). This indicates that the measurement model discriminated adequately between the four factors: \textit{competence, trustworthiness, clarity, engagement}.

To assess the reliability of our measurement model and to ensure that all factors showed satisfactory levels of internal consistency, we examined the composite reliability of all factors. All the values of the composite reliability (CR, Table \ref{tab:cfa}) support the reliability of our measurement instrument, since they were above the cut-off value of .700 \citep{Nunnally1994, Bagozzi1988}. Also all values of Cronbach's $\alpha$ were satisfactory \citep{Cronbach1951}. In conclusion, we assume that our measurement model is able to capture the variables of interest both valid and reliable.

\section{Results}\label{sec:results}

\subsection{Comparison of UI Conditions}
\begin{figure}[ht!]
    \includegraphics[width=1\textwidth]{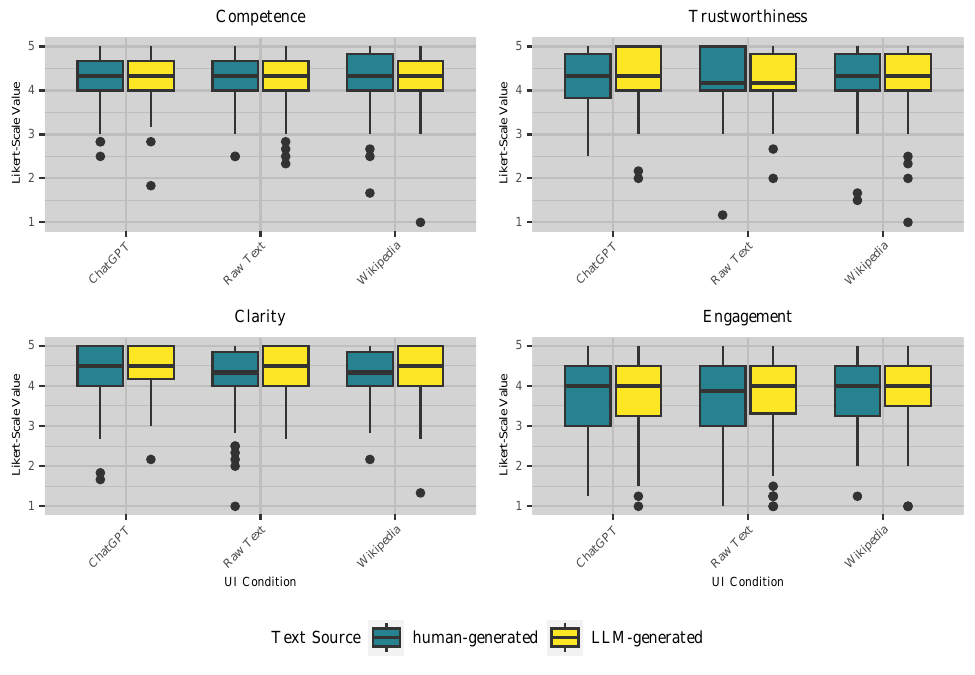}
    \caption{Boxplots of Descriptive Results - Credibility Perceptions per UI Condition and Content Source}
	\label{fig:results_ui}
\end{figure}

We first answer the question whether the UI condition (ChatGPT, Raw Text or Wikipedia) causes participants to have differential credibility perceptions of the presented text excerpts. Figure \ref{fig:results_ui} presents the visual analysis for the four sub constructs of credibility, namely \textit{perceived competence, perceived trustworthiness, perceived clarity and perceived engagement}. The results are presented for each UI condition and text source (human-generated and LLM-generated). 

Firstly, the boxplots show a consistent perception of the presented content as highly credible across all UI conditions. This may be caused by the fact, that we used factual information texts to non-controversial topics without any mistakes. Secondly, if we compare the boxplots for each outcome, we see no substantial differences between the UI conditions. This finding is also supported by the inter-UI-group analysis, utilizing Kruskal-Wallis tests to assess differences across the three UI groups (see Table \ref{tab:results_kw}). Results reveal that the UI groups did not display any substantial variations in all of the assessed dimensions of credibility. The same accounts for the number and rate of errors participants were making after having read the texts in different UI conditions: no differences of statistical significance can be detected. Also the reading time did not differ significantly between UI groups. In summary, these observations collectively suggest that the distinct user interfaces had minimal influence, if any, on users' perceptions of credibility or their comprehension of the content presented.

\begin{table}[ht!]
	\caption{Inter-UI-Group Analysis -- Results of Kruskal-Wallis Tests}
	\centering
	\label{tab:results_kw}
	\footnotesize
	\begin{threeparttable}
	\begin{tabular}{lcccccc}
		\hline\noalign{\smallskip}
	    \multicolumn{7}{c}{Median (SD)}\\
		&ChatGPT (N = 201)& Raw Text (N = 210)& Wikipedia (N = 195) &H & df & $p$\\
		\hline\noalign{\smallskip}
		Competence& 4.250 (.498)& 4.292 (.480) & 4.333 (.524) & 0.002& 2 & .998\textsuperscript{n.s.}\\\noalign{\smallskip}
		Trustworthiness& 4.250 (.579)& 4.250 (.573) & 4.250 (.612)& 0.317 &2&.853\textsuperscript{n.s.}\\\noalign{\smallskip}
		Clarity& 4.500 (.561)& 4.417 (.530)& 4.417 (.545)& 0.855 &2&.652\textsuperscript{n.s.}\\\noalign{\smallskip}
		Engagement& 3.875 (.870) &3.750 (.889)& 3.875 (.749)& 1.572&2&.455\textsuperscript{n.s.}\\\noalign{\smallskip}
		\# Errors\textsuperscript{*}& 0.000 (.585)  & 0.000 (.433)  & 0.000 (.556) & 0.458 & 2 &.795\textsuperscript{n.s.}\\\noalign{\smallskip}
		\% Errors\textsuperscript{**}& 0.000 (.073) & 0.000 (.054) & 0.000 (.069) & 0.458  &2&.795\textsuperscript{n.s.}\\\noalign{\smallskip}
		Reading Time (in sec.)\textsuperscript{a} & 1.976 (1.302) & 1.961 (1.191) & 2.001 (2.027) & 0.132 &2&.935\textsuperscript{n.s.}\\\noalign{\smallskip}
		\hline\noalign{\smallskip}
			\end{tabular}
			\begin{tablenotes}
				\item \scriptsize \textit{Note.} N = 606. \textsuperscript{a} = Normalized with word count per text excerpt. \textsuperscript{n.s.} denotes a non-significant result. \textsuperscript{*} = Overall eight errors were possible since we had two attention checks per text excerpt. \textsuperscript{**} = number of errors / 8.
			\end{tablenotes}
		\end{threeparttable}	
 \end{table}

\subsection{Comparison of Human-Generated versus LLM-Generated Content}
\begin{figure}[ht!]
    \includegraphics[width=1\textwidth]{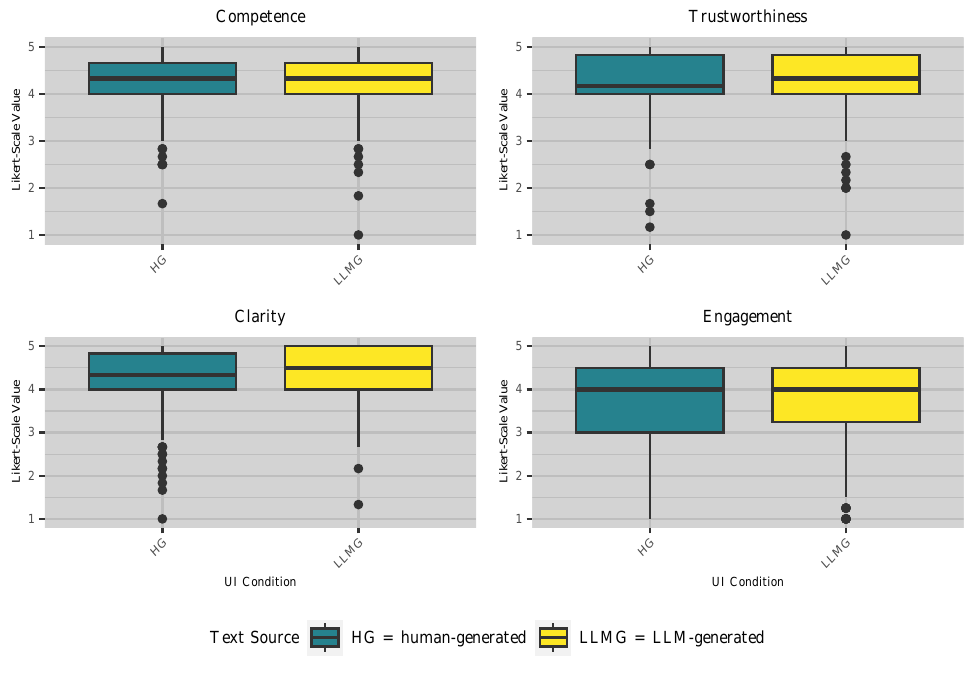}
    \caption{Boxplots of Descriptive Results - Credibility Perceptions per Content Source}
	\label{fig:results_cg_hg}
\end{figure}

Next, we cover the question of whether participants perceive content differently based on its origin -- human-generated or LLM-generated. Figure \ref{fig:results_cg_hg} shows the boxplots for each dimension of credibility. The boxplots reveal that the median values of \textit{perceived competence, trustworthiness, clarity and engagement} do not substantially differ, however clarity and engagement seem to be perceived higher for LLM-generated content. Our post-hoc statistical tests support this visual impression. Table \ref{tab:results_wilcoxon_hg_cg} presents the results of the within-group analysis employing Wilcoxon tests (two-sided). Since every participant both evaluated human and LLM-generated text excerpts (two each) we pooled the participants of each UI condition resulting now in a sample size of 1,212. 

\begin{table}[ht!]
	\caption{Within-Group Analysis -- Results of Wilcoxon Tests (two-sided)}
	\centering
	\label{tab:results_wilcoxon_hg_cg}
	\footnotesize
	\begin{threeparttable}
	\begin{tabular}{lccccc}
		\hline\noalign{\smallskip}
	    \multicolumn{6}{c}{Median (SD)}\\
		&LLM-Generated& Human-Generated  & V & $p$ & Effect Size $r$\\
		\hline\noalign{\smallskip}
		Competence& 4.333 (.541)& 4.333 (.538)  & 46,660.5& .393\textsuperscript{n.s.} & --\\\noalign{\smallskip}
		Trustworthiness& 4.333 (.614)& 4.167 (.623)& 34,823 &.370\textsuperscript{n.s.} & --\\\noalign{\smallskip}
		Clarity& 4.500 (.567)& 4.333 (.564)& 52,912.5 & .000\textsuperscript{***} & 0.261\\\noalign{\smallskip}
		Engagement& 4.000 (.885) & 4.000 (.916)& 53,904.5 & .005\textsuperscript{***} &0.134\\\noalign{\smallskip}
		\# Errors& 0.000 (.086)    & 0.000 (.092) & 2,619.5 & .271\textsuperscript{n.s.} & --\\\noalign{\smallskip}
		\% Errors& 0.000 (.347)  & 0.000 (.370) & 2,619.5  & .271\textsuperscript{n.s.} &--\\\noalign{\smallskip}
		Reading Time (in sec.)\textsuperscript{a} & 0.926 (.950) & 0.977 (1.045) & 79,823.5 & .005\textsuperscript{***} & 0.114\\\noalign{\smallskip}
		\hline\noalign{\smallskip}
			\end{tabular}
			\begin{tablenotes}
				\item \scriptsize \textit{Note.} N = 1,212. \textsuperscript{a} = Normalized with word count per text excerpt. \textsuperscript{n.s.} denotes a non-significant result. \textsuperscript{***} = p $\leq$ .01.
			\end{tablenotes}
		\end{threeparttable}	
 \end{table}
 
 First, we find no substantial difference in terms of \textit{competence} and \textit{trustworthiness}: participants do not perceive these two factors related to source credibility differently. Secondly, however, we find substantial differences with regard to \textit{clarity} and \textit{engagement} which are those two factors that are more strongly related to message credibility itself. Notably, participants perceived the LLM-generated content as clearer compared to human-generated content, as evidenced by a significant $p$-value ($p < 0.01$) and a moderate effect size ($r = 0.261$). Participants perceive the LLM-generated versions of the text excerpts both as more clear and engaging. Similarly, with regards to engagement participants also seem to favor LLM-generated over human-generated content, again with statistical significance ($p < 0.01$), though the effect size was smaller ($r = 0.134$). Finally, while no differences could be detected for the number and rate of errors we find that participants spent significantly less time reading LLM-generated content, indicated by a low p-value ($p < 0.01$) and a small effect size ($r = 0.114$). 
 
\section{Discussion and Implications}\label{sec:discussion}
One of the significant outcomes of this study is the consistent perception of source credibility across UI conditions and content origin. Participants did not substantially differentiate between human-generated and LLM-generated content in terms of \textit{competence} and \textit{trustworthiness}, which also holds for various UI conditions. This suggests that individuals even when they know that content obviously stems from ChatGPT credibility perceptions do not differ with regard to expertise and reliability of the source.

The other relevant finding is that LLM-generated texts produced by ChatGPT are perceived as more clear and captivating. This observation is supported by the fact that participants required notably less reading time for LLM-generated texts. At the same time participants did not demonstrate an increase in errors when answering the attention checks after reading LLM-generated content.

This is an intriguing result: LLM-generated texts seem to outperform human-generated text with regard to clarity and engagement, showing even a higher message credibility than human generated text. At the same time, people do not seem to differentiate between human and LLM-generated texts with regards to source credibility: They perceive the texts as equally competent and trustworthy. The UI has no influence on credibility perceptions at all. So even when people know the origin of the texts they do not doubt the credibility of ChatGPT generated texts. 

Furthermore, public opinion and ongoing media coverage is increasingly recognizing generative AI's proficiency in handling and conveying information. Undoubtedly, LLMs have demonstrated impressive progress in generating high-quality textual content. Nonetheless, it remains a fact that these systems carry an inherent potential for errors, misunderstandings, and even instances of generating content that deviates from reality. Surprisingly, these risks appear to be either underestimated by users or are not substantially influencing their perceptions of credibility.

Since at the core ChatGPT-generated texts are just the result of learning a statistical model on large amounts of data and thus are all but deterministically correct or even curated the presented results are a call for not only reflecting on the use and assimilation of such systems but also the need for regulation and labeling of the origin of information. However, also techniques like watermarking or fingerprinting of AI-generated content \citep{kirchenbauer2023watermark} or using disclaimers \citep{clayton2020real} may not entirely deter individuals from uncritically adopting AI-generated texts. To address this issue, education and increased awareness about how large language models operate must be prioritized. By providing accessible information and training, we can equip people with the tools to better understand and evaluate the risks and uncertainties linked to these systems. This education can help individuals to make informed decisions when and how to use such technologies, ultimately promoting responsible and mindful usage.

\section{Limitations and Future Work}\label{sec:limitations}
While the present study sheds light on the perceived credibility of human-generated and LLM-generated content in various UI versions of text excerpts, there are several limitations that should be acknowledged. However, these limitations suggest directions for future research in the field. 

First, the study primarily focuses on immediate perceptions of credibility and does not delve into the long-term effects that content from different sources might have on recipients. The sleeper effect, as researched by \cite{Hovland1951}, emphasizes how message credibility develops over time especially when message sources are forgotten. Since the current study does not capture long-term effects, it is challenging to ascertain whether such effects play a role in the perceived credibility of human-generated and LLM-generated content, potentially influencing attitudes over extended periods.

Second, the study employed a relatively restricted set of text excerpts, using only four different example texts. This limited scope might not adequately capture the breadth of text types, topics, and sources that individuals encounter in real-world scenarios. Consequently, the findings may not fully generalize to other contexts, thereby constraining the external validity of the study's conclusions. In addition to that, we only concentrated on texts excerpts covering factual information. Future research could explore whether perceptions of credibility differ between human-generated and LLM-generated texts and different UI versions when considering content that presents facts, expresses (political) opinions, or makes evaluative judgments.

Third, although we made efforts to include diverse demographics, the study's sample may not be fully representative of the broader population. Demographic variations, cultural backgrounds, and information consumption habits could all impact the perception of credibility differently. Therefore, the study's findings may not fully encompass the spectrum of perspectives present in society.

\section{Conclusions}\label{sec:conclusion}

In this paper, we studied the credibility perceptions of human-generated and LLM-generated content in different UI settings. We conducted an extensive online survey with more than $600$ participants and asked for their perceived credibility of text excerpts in different UI settings, namely a ChatGPT UI, a Wikipedia UI, and a Raw Text UI, while also manipulating the text's origin which was either human or LLM-generated.

We found that the participants did not substantially differentiate between human-generated and LLM-generated content in terms of \textit{competence} and \textit{trustworthiness}. This also holds regardless of the studied UI. So even if the participants know that the texts are from ChatGPT, they consider them to be as credible as human-generated and curated texts. Furthermore, we found that the texts generated by ChatGPT are perceived as more clear and captivating by the participants than the human-generated texts. 
This perception was further supported by the finding that participants spent less time reading LLM-generated content while achieving comparable comprehension levels.

These results underscore the potential of AI-generated content to enhance the clarity and engagement of information presentation, which could have implications for communication effectiveness and user experiences. However, that users do not distinguish between AI-generated and human-curated texts in terms of \textit{competence} and \textit{trustworthiness}, regardless of origin, poses several risks for future society, as LLM-based systems like ChatGPT are prone to errors and hallucinations, as well as enhance biases present in the training data. Additionally, potential malicious actors can use LLM-based systems to generate fake news or hateful content. This is of particular concern, as according to our study, users do not differentiate between such AI-generated and human-generated content.

Therefore, a labeling requirement for AI-generated content is important and necessary. However, labeling of content alone is not sufficient, as our study showed that users do not critically differentiate between texts, even when their origin is known. Consequently, a stronger development of media competencies for dealing with AI systems and their generated content is needed.

\newpage
\bibliographystyle{agsm}
\bibliography{bibliography}

\appendix

\section{Text Excerpts} \label{Appendix:Text}

\subsection{Human-Generated Text Excerpts (Wikipedia)}

\subsubsection{Acadamy Awards}

The Academy Awards, also known as the Oscars, are awards for artistic and technical merit for the film industry. They are presented annually by the Academy of Motion Picture Arts and Sciences (AMPAS), in recognition of excellence in cinematic achievements as assessed by the Academy's voting membership. The Academy Awards are regarded by many as the most prestigious, significant awards in the entertainment industry in the United States and worldwide. The Oscar statuette depicts a knight rendered in the Art Deco style.

The major award categories are presented during a live televised Hollywood ceremony that is typically held in February or March. It is the oldest worldwide entertainment awards ceremony. The 1st Academy Awards were held in 1929, the second ceremony in 1930 was the first one broadcast by radio, and the 1953 ceremony was the first one televised. It is also the oldest of the four major annual American entertainment awards; its equivalents – the Emmy Awards for television, the Tony Awards for theater, and the Grammy Awards for music – are modeled after the Academy Awards.

\subsubsection{Canada}

Canada is a country in North America. Its ten provinces and three territories extend from the Atlantic Ocean to the Pacific Ocean and northward into the Arctic Ocean, making it the world's second-largest country by total area, with the world's longest coastline. It is characterized by a wide range of both meteorologic and geological regions. The country is sparsely inhabited, with the vast majority residing south of the 55th parallel in urban areas. Canada's capital is Ottawa and its three largest metropolitan areas are Toronto, Montreal, and Vancouver.

Indigenous peoples have continuously inhabited what is now Canada for thousands of years. Beginning in the 16th century, British and French expeditions explored and later settled along the Atlantic coast. As a consequence of various armed conflicts, France ceded nearly all of its colonies in North America in 1763. In 1867, with the union of three British North American colonies through Confederation, Canada was formed as a federal dominion of four provinces. This began an accretion of provinces and territories and a process of increasing autonomy from the United Kingdom, highlighted by the Statute of Westminster, 1931, and culminating in the Canada Act 1982, which severed the vestiges of legal dependence on the Parliament of the United Kingdom.

Canada is a parliamentary liberal democracy and a constitutional monarchy in the Westminster tradition. The country's head of government is the prime minister, who holds office by virtue of their ability to command the confidence of the elected House of Commons and is "called upon" by the governor general, representing the monarch of Canada, the head of state. The country is a Commonwealth realm and is officially bilingual (English and French) in the federal jurisdiction. It is very highly ranked in international measurements of government transparency, quality of life, economic competitiveness, innovation, and education. It is one of the world's most ethnically diverse and multicultural nations, the product of large-scale immigration. Canada's long and complex relationship with the United States has had a significant impact on its history, economy, and culture.

A developed country, Canada has one of the highest nominal per capita income globally and its advanced economy ranks among the largest in the world, relying chiefly upon its abundant natural resources and well-developed international trade networks. Canada is part of several major international and intergovernmental institutions or groupings including the United Nations, NATO, G7, Group of Ten, G20, Organisation for Economic Co-operation and Development (OECD), World Trade Organization (WTO), Commonwealth of Nations, Arctic Council, Organisation internationale de la Francophonie, Asia-Pacific Economic Cooperation forum, and Organization of American States.

\subsubsection{Malware}

Malware (a portmanteau for malicious software) is any software intentionally designed to cause disruption to a computer, server, client, or computer network, leak private information, gain unauthorized access to information or systems, deprive access to information, or which unknowingly interferes with the user's computer security and privacy. Researchers tend to classify malware into one or more sub-types (i.e. computer viruses, worms, Trojan horses, ransomware, spyware, adware, rogue software, wiper and keyloggers).

Malware poses serious problems to individuals and businesses on the Internet. According to Symantec's 2018 Internet Security Threat Report (ISTR), malware variants number has increased to 669,947,865 in 2017, which is twice as many malware variants as in 2016. Cybercrime, which includes malware attacks as well as other crimes committed by computer, was predicted to cost the world economy \$6 trillion USD in 2021, and is increasing at a rate of 15\% per year. Since 2021, malware has been designed to target computer systems that run critical infrastructure such as the electricity distribution network.

The defense strategies against malware differ according to the type of malware but most can be thwarted by installing antivirus software, firewalls, applying regular patches to reduce zero-day attacks, securing networks from intrusion, having regular backups and isolating infected systems. Malware is now being designed to evade antivirus software detection algorithms.

\subsubsection{US Senate}

The United States Senate is the upper chamber of the United States Congress, with the House of Representatives being the lower chamber. Together they compose the national bicameral legislature of the United States.

The composition and powers of the Senate are established by Article One of the United States Constitution. The Senate is composed of senators, each of whom represents a single state in its entirety. Each of the 50 states is equally represented by two senators who serve staggered terms of six years, for a total of 100 senators. From 1789 to 1913, senators were appointed by legislatures of the states they represented. They are now elected by popular vote following the ratification of the Seventeenth Amendment in 1913.

As the upper chamber of Congress, the Senate has several powers of advice and consent which are unique to it. These include the approval of treaties, and the confirmation of Cabinet secretaries, federal judges (including Federal Supreme Court justices), flag officers, regulatory officials, ambassadors, other federal executive officials and federal uniformed officers. If no candidate receives a majority of electors for vice president, the duty falls to the Senate to elect one of the top two recipients of electors for that office. The Senate conducts trials of those impeached by the House. The Senate has typically been considered both a more deliberative and prestigious body than the House of Representatives due to its longer terms, smaller size, and statewide constituencies, which historically led to a more collegial and less partisan atmosphere.

The Senate chamber is located in the north wing of the Capitol Building in Washington, D.C. The vice president of the United States serves as presiding officer and president of the Senate by virtue of that office, despite not being a senator, and has a vote only if the Senate is equally divided. In the vice president's absence, the president pro tempore, who is traditionally the senior member of the party holding a majority of seats, presides over the Senate. In the early 1920s, the practice of majority and minority parties electing their floor leaders began. The Senate's legislative and executive business is managed and scheduled by the Senate majority leader.

\subsection{LLM-Generated Text Excerpts (ChatGPT)}

\subsubsection{Acadamy Awards}

Academy Awards, also known as the Oscars, are a set of annual awards given to recognize outstanding achievements in the film industry. The awards are presented by the Academy of Motion Picture Arts and Sciences, and winners are selected by the organization's voting membership.

The Academy Awards were first presented in 1929 and have since become one of the most prestigious awards in the entertainment industry. The awards are presented in various categories, including Best Picture, Best Director, Best Actor, and Best Actress, among others.

Each year, a number of films are nominated for each category, with winners being announced during a televised ceremony. The awards ceremony is attended by some of the biggest names in the film industry and is watched by millions of people worldwide.

The Academy Awards have been criticized for their lack of diversity in both nominees and winners, and steps have been taken in recent years to increase representation in the industry. Despite this criticism, the awards remain a highly respected honor and a significant milestone for filmmakers and actors alike.

\subsubsection{Canada}

Canada is a country located in North America, bordered by the United States to the south and northwest, the Arctic Ocean to the north, and the Atlantic and Pacific Oceans to the east and west, respectively. With an area of 9.98 million square kilometers, Canada is the second-largest country in the world by total area, after Russia. It is also one of the world's wealthiest countries, with a high standard of living and a well-developed economy.

The population of Canada is around 38 million people, with the majority residing in the southern part of the country. The country is a multicultural society, with English and French being the official languages. The country's capital is Ottawa, while its largest city is Toronto.

Canada is known for its natural beauty, with vast forests, mountains, and lakes. The country is also home to several iconic landmarks, such as Niagara Falls, Banff National Park, and the CN Tower. The country has a diverse economy, with industries such as natural resources, manufacturing, and technology contributing significantly to the country's GDP.

Canada is a parliamentary democracy and a constitutional monarchy, with the British monarch serving as the head of state. The country has a federal system of government, with power divided between the federal government and the provinces and territories. The country has a strong tradition of democratic governance, with a free press, independent judiciary, and a robust civil society.

Canada is known for its progressive social policies, such as universal healthcare, same-sex marriage, and strong environmental protections. The country has a strong social safety net, with programs such as employment insurance, old-age security, and the Canada Pension Plan. The country also has a strong commitment to human rights, with laws protecting freedom of speech, religion, and equality.

Canada is a member of several international organizations, including the United Nations, NATO, and the G7. The country is also a signatory to several international treaties, including the Paris Agreement on climate change and the Comprehensive and Progressive Agreement for Trans-Pacific Partnership.

In conclusion, Canada is a large, diverse, and prosperous country with a strong tradition of democratic governance and progressive social policies. The country is known for its natural beauty, multicultural society, and well-developed economy, and it plays an important role in international affairs.

\subsubsection{Malware}

Malware, short for malicious software, refers to any software or code that is designed to harm or exploit computer systems, networks, or devices. Malware can take many forms, including viruses, worms, trojans, spyware, adware, and ransomware.

Viruses are programs that infect other files or systems, replicating and spreading themselves through networks and devices. Worms, on the other hand, are standalone programs that spread themselves without needing to infect other files. Trojans are disguised as legitimate software but contain harmful code that can cause damage to a computer system or network.

Spyware and adware are types of malware that are designed to monitor a user's activity or display unwanted advertisements. Ransomware, a growing threat, encrypts a user's files and demands a ransom payment in exchange for the decryption key.

Malware can be distributed through a variety of methods, including email attachments, software downloads, and malicious websites. Once installed, malware can cause a wide range of problems, from slowing down a device or network to stealing sensitive information or rendering a system unusable.

Protecting against malware requires using up-to-date antivirus software, regularly updating operating systems and software, and being cautious when downloading and installing new programs. It is also important to back up important data regularly to prevent data loss in the event of a malware attack.

\subsubsection{US Senate}

The United States Senate is one of the two chambers of the United States Congress, the legislative branch of the federal government of the United States. The Senate is composed of 100 Senators, two from each state, who are elected to six-year terms.

The Senate has several important powers and responsibilities. It shares with the House of Representatives the power to make laws, but also has the power to approve or reject presidential appointments, including those of federal judges and executive branch officials. The Senate is also responsible for conducting impeachment trials of federal officials, including the President of the United States.

The Senate is led by the Vice President of the United States, who serves as its President, but can only vote in the event of a tie. In the absence of the Vice President, the President pro tempore, who is usually the most senior member of the majority party, presides over the Senate.

The Senate has a number of committees that oversee various aspects of federal policy and conduct investigations. These committees are often chaired by Senators who have particular expertise or interest in the issues they address. Some of the most powerful committees in the Senate include the Appropriations Committee, the Armed Services Committee, and the Finance Committee.

The Senate is known for its unique traditions and customs. Senators are often addressed as "Senator" followed by their last name, and are expected to show respect for their colleagues by not using personal attacks or insults during debates. The Senate also has a number of procedural rules that are designed to ensure that all Senators have an opportunity to participate in debates and that the minority party has some ability to influence legislation.

In recent years, the Senate has been characterized by increasing levels of partisanship and gridlock, with many bills and nominations stalled due to political disagreements. However, the Senate remains an important institution in American democracy and plays a vital role in shaping the direction of the country.

\section{Comprehension Questions - Attention Checks} \label{Appendix:attention_check}
The following section shows the comprehension questions per text excerpt, which serve as attention checks. Each attention check begins with the question, followed by the possible answers in italics, with the correct answer marked in \textbf{bold} font. Attention checks did not differ between the UI condition or between human-generated and LLM-generated text excerpts.

\subsection{Academy Awards}

\begin{itemize}
    \item In which decade were the Academy Awards first presented? \textit{1910s / \textbf{1920s} / 1930s / 1940s}
    \item By what name are the Academy Awards better known? \textit{Golden Globe Awards / Emmy Awards / Tony Awards / \textbf{Oscars}}
\end{itemize}

\subsection{Canada}

\begin{itemize}
    \item What is the capital of Canada? \textit{Montreal / \textbf{Ottawa} / Toronto / Vancouver}
    \item Which of the following international institutions is not mentioned in the text? \textit{United Nations / NATO / \textbf{World Health Organization (WHO)} / G7}
\end{itemize}

\subsection{Malware}

\begin{itemize}
    \item What is a subtype of malware? \textit{Freeware / Shareware / \textbf{Ransomware} / Software}
    \item What method to protect yourself against malware is NOT mentioned in the text? \textit{Antivirus software /  Regular backups / \textbf{Specific internet browsers} / Regular updates}
\end{itemize}

\subsection{US Senate}

\begin{itemize}
    \item How many senators per state are in the Senate? \textit{\textbf{2} / 3 / 4 / 5}
    \item Who is the presiding officer and president of the Senate? \textit{The president of the United States  / \textbf{The vice president of the United States} / The speaker of the House of Representatives / The United States Secretary of State}
\end{itemize}

\end{document}